\PassOptionsToPackage{dvipsnames}{xcolor}
\documentclass[twocolumn]{aastex62}
\usepackage{graphicx} \usepackage{xcolor}
\usepackage{array}
\usepackage{multirow}

\usepackage{fdsymbol}
\newcommand\psj{\ref@jnl{PSJ}}%

\begin{document}

\title{Hydrolyzed Hazes on Water-rich Exoplanets: Optical Constants and Detectability}

\author[0009-0007-6910-6347]{Cara Pesciotta}
\affil{Department of Earth and Planetary Sciences, Johns Hopkins University, Baltimore, MD, USA}

\author[0000-0003-4596-0702]{Sarah M. H{\"o}rst}
\affil{Department of Earth and Planetary Sciences, Johns Hopkins University, Baltimore, MD, USA}

\author[0000-0002-4072-181X]{Michael J. Radke}
\affil{Department of Earth and Planetary Sciences, Johns Hopkins University, Baltimore, MD, USA}

\author[0000-0002-6721-3284]{Sarah E. Moran}
\altaffiliation{NHFP Sagan Fellow}
\affil{NASA Goddard Space Flight Center, Greenbelt, MD, USA}
\affil{Space Telescope Science Institute, Baltimore, MD, USA}

\author[0000-0002-6694-0965]{Chao He}
\affil{School of Earth and Space Sciences, University of Science and Technology of China, Hefei, China}

\author[0000-0001-7273-1898]{V{\'e}ronique Vuitton}
\affil{Univ. Grenoble Alpes, CNRS, IPAG, 38000 Grenoble, France}

\begin{abstract}
Observations of temperate sub-Neptunes suggest active chemical environments, finding evidence of both water vapor and photochemical hazes in their atmospheres. Hazes formed in water-rich atmospheres are chemically complex, containing molecules relevant to prebiotic chemistry, and their strong optical opacity obscures sought-after gaseous molecular absorption features. While many studies have investigated haze formation and properties across diverse atmospheric conditions, little is known about the evolution of these hazes in their environment once formed. In particular, interactions with water can drive hydrolysis reactions that alter haze composition and optical behavior, affecting our interpretations of habitability and observational spectroscopy. Here, we perform hydrolysis experiments on haze analogs of temperate water-rich exoplanets and measure their optical properties. Transmittance measurements from 0.4 to 28.5 $\mu$m reveal changes in key functional groups after hydrolysis, along with an overall increase in sample absorbance. We report the derived optical constants for use in observational and modeling studies. Through synthetic atmospheric spectra, we demonstrate the need for physically informed haze optical properties in models, consistent with expected planetary conditions. The increased absorptivity and high imaginary refractive index of hydrolyzed hazes almost completely flatten features in model spectra, presenting critical consequences for atmospheric characterization of water-rich sub-Neptunes.

\vspace{1cm}
\end{abstract}

\section{Introduction}
The majority of the over 6000 confirmed exoplanets fall into the categories of super-Earths and sub-Neptunes, together defined as having radii between 1.5 and 4 R$_E$. These mysterious planets have no Solar System analogs, and their bulk densities alone are often degenerate, consistent with either a silicate-iron core with a hydrogen-rich atmosphere or an icy core with a water-rich atmosphere \citep{Adams2008,Valencia2013,Fulton2017,Luque2022}. Ground- and space-based observations can help break this degeneracy, identifying atmospheric absorption features to reveal the composition. Tentative detections of water have been reported in the atmospheres of HAT-P-11b \citep[T$_{eq}\sim$800 K;][]{Fraine2014,Chachan2019,Basilicta2024}, HAT-P-26b \citep[T$_{eq}\sim$1000 K;][]{Wakeford2017,Gressier2025}, TOI-270d \citep[T$_{eq}\sim$350 K;][]{Benneke2024}, and GJ 3470b \citep[T$_{eq}\sim$600 K;][]{Benneke2019b,Beatty2024}, and, contentiously, a water-rich reservoir has been inferred for K2-18b \citep[T$_{eq}\sim$250 K;][]{Benneke2019,Madhusudhan2023,Schmidt2025}. All reported equilibrium temperatures, T$_{eq}$, for the aforementioned planets assume an albedo around $\sim$0.3, which is typical for exoplanet literature. Many exoplanet observations, however, yield muted or even flat transmission spectra that provide evidence of high-altitude aerosols such as clouds and/or hazes. 

Photochemical hazes are a widespread feature of planetary atmospheres, present close to home on bodies such as Titan \citep{Cable2012} and Venus \citep{Titov2018} and potentially extending to distant exoplanets including the super-Earth HD 97658b \citep{Knutson2014b} and warm Neptunes GJ 1214b \citep{Kreidberg2014,Kempton2023,Gao2023,Schlawin2024,Ohno2025}, GJ 436b \citep{Knutson2014,Lothringer2018}, and GJ 3470b \citep{Dragomir2015}. Abundant haze formation is expected on temperate exoplanets \citep[$<$1000 K;][]{Gao2020,Morley2013,Horst2018}, with population statistics indicating temperature-dependent formation therein \citep{Estrela2022,Brande2024}. Hazes affect a planet's temperature structure, atmospheric dynamics, cloud formation, and even supply organic material to the surface \citep[e.g.,][]{Marley2013,Lavvas2021,Steinrueck2025}. Hazes are also reservoirs of prebiotic chemistry, comprised of long hydrocarbon chains that may build into amino acids, nucleobases, and sugars \citep[e.g.,][]{Horst2012,Moran2020}. Given the prebiotic relevance, the presence of hazes on water-rich worlds makes them compelling targets in the search for life on other planets. 

Beyond this chemical significance, the muted spectral features and large scattering slopes that denote photochemical hazes in spectra obscure other molecular signatures. Studies have shown that aerosols may dampen the 1.4 $\mu$m water feature by 50-70\% as a result of their competing opacity \citep{Gao2021}. Characterizing haze physical, optical, and chemical properties is therefore paramount to analyzing and understanding the planetary environment. 

Models and laboratory studies of hazes demonstrate that particles formed under different atmospheric conditions exhibit distinct properties. For haze formation simulated in temperate conditions, the initial atmospheric composition strongly influences the resulting haze. For instance, hazes formed under H$_2$O- or CO$_2$-rich conditions incorporate significantly more oxygen than Titan-like hazes produced from N$_2$ and CH$_4$ \citep{Moran2020}. These chemical differences lead to measurable changes in their scattering slopes and absorption features, which will in turn effect the interpretation of transmission and emission spectra \citep{Gavilan2018,Corrales2023,He2024,Li2025}. This highlights the importance of adopting optical properties tailored to the conditions on the planet in question.

While studies have characterized several properties that govern haze microphysics (e.g., optical properties, \citealt{Gavilan2018,Corrales2023,He2024,Drant2024}; surface energy, \citealt{Yu2021}; composition, \citealt{Moran2020,Wang2025}), the fate of haze particles as they evolve in their environment is largely unexplored. On temperate water-rich planets, hazes are likely to react with ambient water both in the atmosphere and on the surface. Haze particles are favorable cloud condensation nuclei (CCN; \citealt{Yu2021}), allowing reactions to occur at the interface between the particles and condensed water in the atmosphere \citep{Maillard2023}. Hazes also readily deposit on the surface through wet or dry deposition \citep{Yu2021}, where they may meet liquid water lakes or oceans \citep{Seager2013,Arney2016,Loftus2019}. We expect haze particles to experience physical and chemical changes as a result of these interactions that consequently alter their optical properties. 

In this work, we perform the first laboratory hydrolysis experiments on exoplanet haze analogs. We measure their transmittance from 0.4 to 28.5 $\mu$m (25000 to 350 cm$^{-1}$), covering visible Hubble wavelengths and all of the JWST wavelength range, and derive their optical constants, providing critical inputs for interpreting observations of temperate water-rich exoplanets. We then model a GJ 1214b-like planet under different haze scenarios to examine hydrolyzed haze effects on transmission and emission spectra.

\section{Methods}

\subsection{Haze Analog Production}
\begin{figure*}
    \centering
    \includegraphics[width=0.8\textwidth]{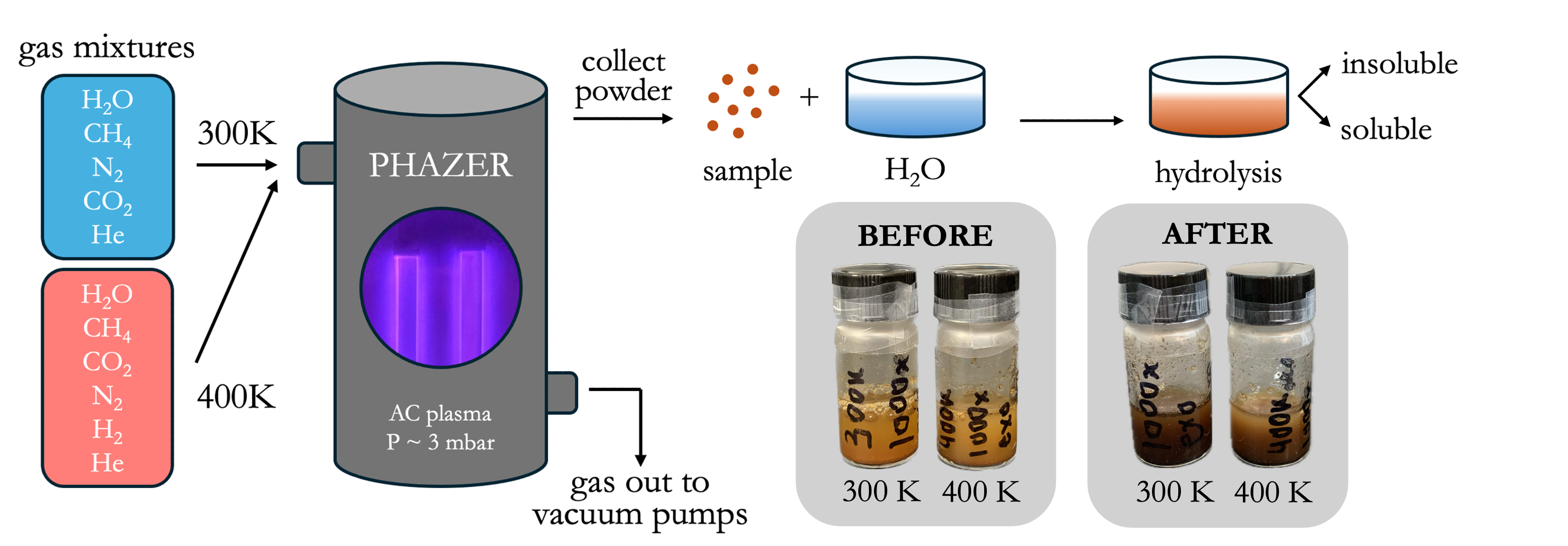}
    \caption{Simplified schematic of the experimental setup. The 1000x solar metallicity gas mixtures are heated to their respective temperatures and flowed through the PHAZER reaction chamber. The collected powder is mixed with water; gray panels display images of the hydrolysis solutions before and after three weeks, demonstrating chemical evolution by eye. The solutions are then separated into insoluble and soluble fractions after hydrolysis.}
    \label{fig:setup}
\end{figure*}

Haze analogs are produced using the Planetary Haze Research (PHAZER) chamber at the Johns Hopkins University \citep{He2017}, designed to investigate photochemistry in a range of planetary atmospheres. The experimental setup is described in detail in \cite{He2017}, and experimental procedures for Titan-like atmospheres \citep{He2022} and possible exoplanet atmospheres \citep{Horst2018} are also well-documented. Figure \ref{fig:setup} illustrates a summary of the setup and procedure with relevant aspects described below.

Table \ref{tab:mixingratio} specifies gas mixing ratios for the targeted atmospheric conditions. {The water-rich exoplanet mixing ratios are determined through equilibrium chemistry calculations of 1000x solar metallicity gases at the two chosen temperatures, 300 and 400 K, using the NASA CEA code \citep{Gordon1994,Moses2013,Horst2018}. At equilibrium, the resulting gas mixtures are no longer at a scaled solar C/O ratio due to condensation changing the elemental ratios.} These mixtures are heated to 300 and 400 K using a custom heating coil. Gas flows through the stainless steel chamber at a constant rate of 10 standard cm$^3$ minute$^{-1}$, maintaining a constant pressure of 2 Torr ($\sim$3 mbar) in the chamber. The gas is exposed to cold plasma discharge for approximately 3 s, producing solid photolysis products. The plasma energy ranges from 1.5 to 18.5 eV with an energy density of 170 W m$^{-2}$ \citep{Pearce2022}. Resulting particles are collected from the chamber and stored in a dry N$_2$ glove box.

\begin{table}[h!]
\centering
\caption{Initial gas mixtures for haze analog production}
\begin{tabular}{l c c}
    \hline \hline
    \multirow{2}{*}{Planet} & \multirow{2}{*}{Mixing Ratio} & Production \\
     & & Rate (mg h$^{-1}$) \\
    \hline
    300 K & 66.0\% H$_2$O & 10.43 \\
    Water-rich & 6.6\% CH$_4$ & \\
    Exoplanet & 6.5\% N$_2$ & \\
     & 4.9\% CO$_2$ & \\
     & 16.0\% He & \\
    \hline
    400 K & 56.0\% H$_2$O & 10.00 \\
    Water-rich & 11.0\% CH$_4$ & \\
    Exoplanet & 10.0\% CO$_2$ & \\
     & 6.4\% N$_2$ & \\
     & 1.9\% H$_2$ & \\
     & 14.7\% He & \\
\end{tabular}
\label{tab:mixingratio}
\end{table}

\subsection{Haze Hydrolysis}
Haze analog samples are then exposed to water for hydrolysis experiments. Procedures are adapted from previous Titan haze hydrolysis experiments \citep{Neish2008,Ramirez2010,Poch2012,Cleaves2014}. For each haze analog, 30 mg of sample is added to 2 mL HPLC (High-Performance Liquid Chromatography) grade water and sealed in a glass vial. A small amount of atmosphere exists inside the vial, however the oxygen content of the atmosphere is negligible compared to that of water. The mixture sits for three weeks at room temperature (294 K), ensuring sufficient reaction time according to the prior experiments. Vials are periodically shaken ($\sim$once per day) to increase the surface area of haze sample reacting with water. After three weeks, the resulting hydrolyzed haze is split into liquid soluble and solid insoluble fractions using a centrifuge at 4200 RPM for 10 minutes. 

Water and any volatile products in the soluble fraction evaporate away, leaving behind a sticky substance. Any water remaining in the insoluble component also evaporates away. We note that the 400 K water-rich exoplanet sample did not produce enough of the soluble component for analysis. The majority of mass loss is likely due to the material properties-- a mix of highly volatile products and refractory reactants. Material loss due to entrainment (fine solid particles carried away during water evaporation) or air currents in the fume hood is possible but likely minimal. If any material was removed during drying, the derived optical constants may slightly underestimate the total absorption of the samples or alter the relative amplitudes of functional group features. 

\subsection{Vacuum FTIR Measurements}
Transmittance spectra of the hydrolyzed haze are measured with a Vertex 70v Fourier transform infrared spectrometer (FTIR; Bruker Optics). The spectrometer is under vacuum ($<$0.2 mbar), reducing atmospheric spectral features, and kept at room temperature (294 K). The covered spectral range is 0.4 to 28.5 µm (25000 to 350 cm$^{-1}$) with a 2 cm$^{-1}$ resolution. 

The three hydrolyzed haze samples (insoluble fractions of both planet analogs, soluble fraction of the 300 K planet analog) are processed into potassium bromide (KBr) pellets to enable measurement. KBr, transparent at the measured wavelengths, acts as a binding agent and dilutant so the sample concentration can be optimized for transmittance measurements. Steps for making a quality KBr pellet are described in \cite{He2024}. The pressed pellets are 13 mm in diameter with effective thickness
\begin{equation}
    d = \frac{m}{\pi r^2 \rho}
    \label{eq:d}
\end{equation}
where $m$ is the mass of the hydrolyzed haze analog in the pellet, $r$ is the pellet radius (6.5 mm), and $\rho$ is the density of the hydrolyzed haze. Since density measurements require a relatively large amount of sample, we use the densities of unhydrolyzed haze analogs: 1.328 g cm$^{-3}$ for the 300 K water-rich exoplanet and 1.262 g cm$^{-3}$ for the 400 K water-rich exoplanet \citep[measurement uncertainties $<$1\%;][]{He2024}. A discussion regarding the sensitivity of our results to the assumed densities can be found in Appendix \ref{sec:sensitivity}. Concentration, mass, and effective thickness of our pellets are reported in Table \ref{tab:pellets}. Physical properties for unhydrolyzed haze pellets (measured in \citealt{He2024}) are also included in this table since the spectra are reprocessed in this work for fair comparison. 

\begin{table*}[t]
\centering
\caption{Physical properties of KBr pellets for FTIR measurements}
\begin{tabular}{l l c c c}
    \hline \hline
    Planet & Sample & Concentration & Mass (mg) & Effective Thickness ($\mu$m) \\
    \hline
    300 K water-rich exoplanet & Insoluble & 0.48\% & 206.4 & 5.6 \\
    & Soluble & 0.12\% & 200.3 & 1.4 \\
    & Unhydrolyzed & 0.38\% & 185.3 & 4.0 \\
    \hline
    400 K water-rich exoplanet & Insoluble & 0.49\% & 204.2 & 6.0 \\
    & Unhydrolyzed & 0.44\% & 193.2 & 5.1 \\
\end{tabular}
\label{tab:pellets}
\end{table*}

KBr pellets are placed in a sample holder normal to the light source for measurement. To achieve the full broadband coverage (0.4 to 28.5 µm), multiple detectors and beam splitters are required. A silicon diode detector and quartz beam splitter are used for 0.4 to 1.11 µm (25000 to 9000 cm$^{-1}$); a DLaTGS detector and quartz beam splitter are used for 0.83 to 2.86 µm (12000 to 3500 cm$^{-1}$); and a DLaTGS detector and KBr beam splitter are used for 1.25 to 28.6 µm (8000 to 350 cm$^{-1}$). The FTIR is set to a 1 mm aperture. Obtained spectra are averaged over 500 scans with a resolution of 2 cm$^{-1}$. The empty sample holder as well as a pure KBr pellet are also measured for reference.

These datasets are processed to eliminate instrumental bias and stitched together into one continuous spectrum. First, the transmittance is obtained by dividing the sample measurement by the reference measurements. Then the peak from the laser at 633 nm is removed. If interference fringes are seen in the spectrum (typically in the mid-IR range, 1.11 to 28.6 µm), they are removed using the method described in \cite{Neri1987}. The fringe-removed transmittance is given by
\begin{equation}
    F(x_n) = \frac{2G(x_n) + G(x_{n+m}) + G(x_{n-m})}{4}
\end{equation}
where $F(x_n)$ is the fringe-removed transmittance at wavelength $x_n$, $G(x_n)$ is the original transmittance at $x_n$, and $G(x_{n+m})$ and $G(x_{n-m})$ are original transmittance values at shifted wavelengths. The wavelengths are shifted by $m$, where $2m$ is the integer number of points contained in the interval $d$, which is the average fringe spacing. The fringe spacing ($d$) is determined for each individual spectrum. Lastly, the three spectral ranges are stitched together, taking advantage of overlapping data to align intensities and eliminate noise at detector limits. These data-processing steps minimize undesired artifacts resulting from sample geometry and instrumental bias. 

\subsection{Optical Constants Derivation}
The optical constants (the real and imaginary parts of the complex refractive index, $N=n+ik$) can be determined using transmittance spectra of the samples. Absorbance ($A$) is given by
\begin{equation}
    A = -\ln(T)
\end{equation}
where T is the transmittance as obtained by FTIR measurements. From the Beer-Lambert law, the absorption coefficient ($\alpha$) is equal to
\begin{equation}
    \alpha (\nu) = \frac{A}{d} = \frac{-ln(T)}{d}
\end{equation}
where $\nu$ is the wavenumber and $d$ is the effective thickness as calculated from Equation \ref{eq:d}. Then, the imaginary refractive index or extinction coefficient ($k$) can be calculated by
\begin{equation}
    k (\nu) = \frac{\alpha (\nu)}{4 \pi \nu} = \frac{-ln(T)}{4 \pi \nu d}
\end{equation}
The real refractive index ($n$) is determined using the extinction coefficient, based on the subtractive Kramers-Kronig (SKK) relation \citep{Wood1982,Toon1994,Imanaka2012} between $n$ and $k$
\begin{equation}
    \label{eq:rri}
    n (\nu) = n_0 + \frac{2(\nu^2 - \nu_0^2)}{pi} P \int_0^\infty \frac{\nu' k(\nu')}{(\nu'^2 - \nu^2) (\nu'^2 - \nu_0^2)} d\nu'
\end{equation}
where $n_0$ is the real refractive index at $\nu_0$ and $P$ is the Cauchy principal value. The values $n_0$ and $\nu_0$ are anchor points to reduce uncertainty in numerical integration \citep{Toon1994,Imanaka2012}. We assume an anchor point from the unhydrolyzed haze optical constant derivation in \cite{He2024}.  Determining the Cauchy principal value $P$ with direct numerical integration is nontrivial and computationally intensive, so we utilize the Maclaurin approximation method due to its accuracy and computational speed \citep{Ohta1988}. Derived $k$ values are used from 350 to 25000 cm$^{-1}$ and $k$ is assumed constant outside of this range. This is a valid assumption as long as no large local absorption peaks exist outside the range. Appendix \ref{sec:sensitivity} demonstrates that our results are robust to assumptions in this calculation including the choice of anchor point and treatment of $k$ outside of the measured range.

Uncertainty in the optical constant calculation is dominated by the transmittance measurements, the effective thickness calculation, and assumptions made throughout. There is systematic uncertainty from the spectrometer, particularly at the edges of the detectors and in the mid-IR range. Overlapping detection ranges reduce error, and we remove interference fringes as described above. For the effective thickness, measurement uncertainty in the sample mass and density is taken into account. We also consider error introduced by assuming a particle density, anchor point, and k values outside of the measured range. Appendix \ref{sec:sensitivity} discusses these assumptions and uncertainties in detail.

\subsection{Atmospheric Spectra Models}
To examine the effect of various hazes in atmospheric spectra, we use the open-source aerosol modeling code \textit{Virga} \citep{Batalha2025b} and open-source radiative transfer suite \textit{PICASO} \citep{Batalha2019} to simulate transmission and emission spectra of a representative planet. We simulate a GJ 1214b-like planet, using the stellar and planetary parameters reported in \cite{Mahajan2024} with an isothermal pressure-temperature profile at the planet's equilibrium temperature of 600 K for simplicity. The atmospheric composition is comprised of the mixing ratios used in our 400 K water-rich exoplanet experiment.

The optical constants from our haze analog experiments are fed into \textit{Virga}, where the program calculates the wavelength-dependent Mie coefficients from the provided $n$ and $k$ values. The calculations use the particle size distribution of the 400 K exoplanet haze analog (40--90 nm, D$_{mean}$ = 57.9 nm) from \cite{He2018}. It is important to note that \textit{Virga} assumes a spherical particle, whereas actual photochemical hazes may be aggregate particles of varying sizes and porosity that affect scattering differently \citep{Adams2019,Vahidinia2024}. We define a homogeneous haze layer from 0.1 bar to 0.1 $\mu$bar with a number density of $\sim$32 cm$^{-3}$, consistent with the extent of Titan's detached haze layer \citep{Lavvas2009} as well as theoretical modelling of exoplanet haze formation \citep{Kawashima2018}. This haze profile is used in \textit{PICASO} to generate synthetic spectra. We use \textit{PICASO}'s built-in opacity database (version 2, \citealt{Batalha2020}) including the species CH$_4$, CO, CO$_2$, Cs, H$_2$O, H$_2$S, K, Li, N$_2$O, NH$_3$, Na, O$_2$, O$_3$, PH$_3$, Rb, TiO, and VO. The opacities span 0.3 to 14 $\mu$m and are resampled to R=$10^4$ from an original R=$10^6$. We generate transmission and emission spectra of the GJ 1214b-like planet with a clear sky, Titan-like haze (using the optical properties of \citealt{Khare1984}), and water-rich haze in addition to the soluble and insoluble fractions of the hydrolyzed haze.

\section{Results and Discussion} 

\subsection{Physical Properties of Hydrolyzed Haze}
\label{sec:solubility}
The haze analogs change in color during hydrolysis, demonstrating chemical processing by eye (see Figure \ref{fig:setup}). The original haze analogs for the 300 and 400 K atmospheres appear a similar gold color and maintain a similar bright orange color when first mixed in water. After 3 weeks, the colors evolve and differentiate. The 300 K haze-water mixture is dark brown, while the 400 K haze-water mixture is a lighter brown. This difference is likely explained by differing chemical compositions of the samples that lead to less absorption in the 400 K sample at visible wavelengths.

\begin{table}[t]
\centering
\caption{Solubility results for water-rich exoplanet hazes.}
\begin{tabular}{l c c}
    \hline \hline
    Quantity & 300 K Exo & 400 K Exo \\
    \hline
    Starting mass (mg) & 32.8 & 32.5 \\
    Collected masses & & \\
    \hspace{5pt} Insoluble (mg) & 4.8 & 4.9 \\
    \hspace{5pt} Soluble (mg) & 1.2 & -- \\
    Solubility (mg/mL) & 13.4 & 13.8 \\
\end{tabular}
\label{tab:solubility}
\end{table}

When separated into soluble and insoluble fractions and dried, the soluble fraction is sticky and viscous, while the insoluble fraction is solid and appears powdery. Table \ref{tab:solubility} summarizes the starting mass, collected soluble and insoluble fraction masses, and resulting upper limit solubility for our two haze scenarios. The derived solubilities are upper limits due to limitations in our ability to collect and measure the tacky soluble fraction as well as any potential losses during the drying process. We include the dried soluble fraction mass in the solubility calculation since this material would remain in a semisolid phase at the relevant temperatures. 

The solubilities are relatively large at 13.4 and 13.8 mg/mL for the 300 and 400 K samples, respectively, with each sample losing a large amount of solid mass in both the soluble and insoluble fractions. This suggests that hydrolysis reactions efficiently produce liquid products, many of which are highly volatile and evaporate away in the drying process. In a planetary atmosphere, these volatiles may themselves be condensable and form new cloud layers. This is particularly evident in the 400 K exoplanet, where we were unable to collect any solid soluble sample. It remains uncertain whether increasing the starting mass would yield a larger soluble mass, or if this haze composition simply does not produce refractory soluble molecules upon hydrolysis. Future work should aim to characterize the liquid hydrolysis products as well as perform a full quantitative solubility study. 

Aerosol solubility is an important parameter in cloud microphysics. For a cloud droplet to form, the relative humidity of a condensable vapor must exceed 100\% (supersaturation). The degree of supersaturation required for a cloud droplet to form on a CCN is determined by the K{\"o}hler Equation, relating CCN curvature and solute effects to supersaturation \citep{Kohler1936}. The equation dictates that CCN with a small contact angle and high solubility in the condensate require a lower supersaturation to activate \citep{Mircea2002}. Previous measurements of the exoplanet analogs studied here \citep{Yu2021} find that the 300 K water-rich exoplanet haze analog indeed has a small contact angle with water ($\theta$=33$^{\circ}$), while the 400 K analog has a large contact angle ($\theta$=65$^{\circ}$). The high solubilities found in this study, however, suggest that both analogs are good CCN regardless of contact angle (solubilites $>$10 mg/mL, according to \citealt{Raymond2002}).

With such a high water solubility, hazes formed in environments similar to our experiments may readily promote water cloud formation. Clouds, in addition to hazes, greatly mute spectral features and are difficult to distinguish from an airless planet \citep{Komacek2020}. Furthering our understanding of these aerosols, including when and where they are likely to occur, is paramount for the future of temperate planet observations. The solubilities reported here can be used for first-order calculations estimating supersaturation and cloud formation in sub-Neptune atmospheres.

\newcommand{\br}{\textcolor{RawSienna}{$\medblacksquare$}}
\newcommand{\ora}{\textcolor{Orange}{$\medblackdiamond$}}
\newcommand{\gr}{\textcolor{OliveGreen}{$\medblackcircle$}}
\newcommand{\whs}{\textcolor{White}{$\medblacksquare$}}
\newcommand{\whd}{\textcolor{White}{$\medblackdiamond$}}
\newcommand{\whc}{\textcolor{White}{$\medblackcircle$}}
\begin{table*}[hp]
\centering
\caption{Functional group assignments for the absorption features in Figure \ref{fig:transmittance}. Features present in the insoluble fraction, soluble fraction, and unhydrolyzed samples are indicated by brown squares (\br), orange diamonds (\ora), and green circles (\gr), respectively.}
\begin{tabular}{c l l c c c}
    \hline \hline
    Wavenumber (cm$^{-1}$) & Functional Group & Intensity & 300K Exo & 400K Exo \\
    \hline
    3500-2500 & O-H stretching, alcohols and carboxylic acids & strong, broad & \br\ora\gr & \br\gr \\
    3413 & free N-H stretching & medium & \ora & -- \\
    3360-3323 & asymmetrical N-H stretching & medium & \br\ora\gr & \br\gr \\
    3300-3030 & N-H stretching, ammonium ions & strong, broad & \br\ora\gr & \br\gr \\
    3188 & symmetrical N-H stretching, amide & medium & \br\ora\gr & \br\gr \\
    3000-2800 & N-H stretching, amine salts & strong, broad & \br\whd\gr & \br\gr \\
    2967 & asymmetrical C-H stretching, methyl group & strong, sharp & \br\ora\gr & \br\gr \\
    2935 & asymmetrical C-H stretching, methylene group & strong, sharp & \br\ora\gr & \br\gr \\
    2877 & symmetrical C-H stretching, methyl group & strong, sharp & \br\ora\gr & \br\gr \\
    2242 & C$\equiv$N stretching & medium & \br\ora\gr & \br\gr \\
     & C$\equiv$C stretching & weak & & \\
    2170 & conjugated C$\equiv$N stretching & strong & \br\ora\gr & \br\gr \\
    2147-2136 & N=C=N stretching & medium & \whs\ora\whc & -- \\
    2097 & C$\equiv$N stretching & medium & \whs\whd\gr & -- \\
     & N=C=N stretching & medium & & \\
    2000-1800 & overtone bands, aromatic rings & weak & \br\ora\gr & \br\gr \\
    1660 & C=O stretching & strong & \br\ora\gr & \br\gr \\
     & N=O stretching & strong & & \\
     & C=C stretching & medium & & \\ 1630 & C=O stretching & strong & \whs\ora\gr & \whs\gr \\
     & C=C stretching & strong & & \\
     & N-H bending & medium & & \\
    1550 & N-H bending & strong & \br\whd\gr & \whs\gr \\
     & N=O stretching & medium & & \\ 1456 & asymmetrical C-H bending, methyl group & strong, sharp & \br\ora\gr & \br\gr \\
     & C-H scissoring, methylene group & strong, sharp & & \\
     & C-C ring stretch & strong & & \\
     & N=O stretching & strong & & \\
    1383 & symmetrical C-H bending, methyl group & strong, sharp & \br\ora\gr & \br\gr \\
     & O-H bending & medium & & \\ 1340 & C-H twisting, methylene group & weak & \whs\ora\whc & -- \\
    1300-1100 & C-C-C stretching & medium & \br\ora\gr & \br\gr \\
     & C-O stretching & medium & & \\
     & C-C(=O)-C bending & weak & & \\
     & C-N stretching & weak & & \\
    1200-800 & C-C stretching & weak, broad & \br\ora\gr & \br\gr \\
    764 & out-of-plane C-H bending, aromatic rings & medium & \br\ora\gr & \br\gr \\
     & out-of-plane N-H wagging & medium, broad & & \\
     & N-O stretching, nitrile & medium & & \\
    704 & C-H rocking, methylene group & medium & \whs\whd\gr & \whs\gr \\
     & out-of-plane N-H wagging & medium, broad & & \\
    595 & out-of-plane C-H bending, aromatic rings & medium & \br\ora\gr & \br\gr \\ 
    \vspace{1pt} \\
    \hline
\end{tabular}
\label{tab:funcgroups}
\vspace{2mm}
\parbox{\textwidth}{\raggedright
  \small {Note.} Functional group assignments based on \cite{Silverstein2017}.}
\end{table*}

\begin{figure*}[th!]
    \centering
    \includegraphics[width=0.9\textwidth]{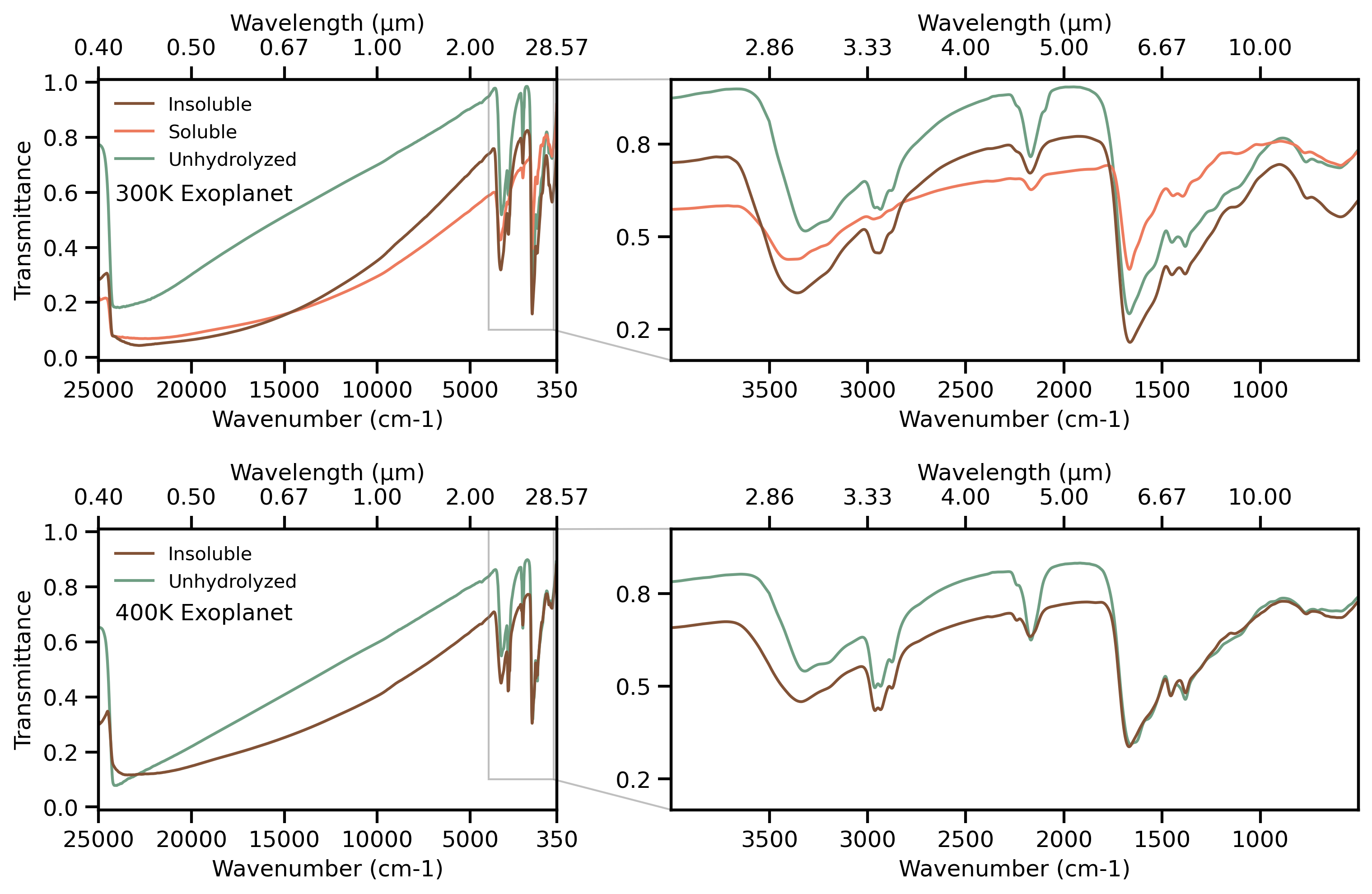}
    \caption{Transmittance spectra of the insoluble (pellet concentrations $\sim$0.5\%) and soluble ($\sim$0.1\%) portions of hydrolyzed haze analogs, and the unhydrolyzed (0.38\% and 0.44\%) sample for comparison. Left panels are the full spectral range, and right panels are the mid-IR range. Top: 300 K water-rich exoplanet; Bottom: 400 K water-rich exoplanet.}
    \label{fig:transmittance}
\end{figure*}

\subsection{Transmittance of Hydrolyzed Haze}
\begin{figure*}[th!]
    \centering
    \includegraphics[width=0.8\textwidth]{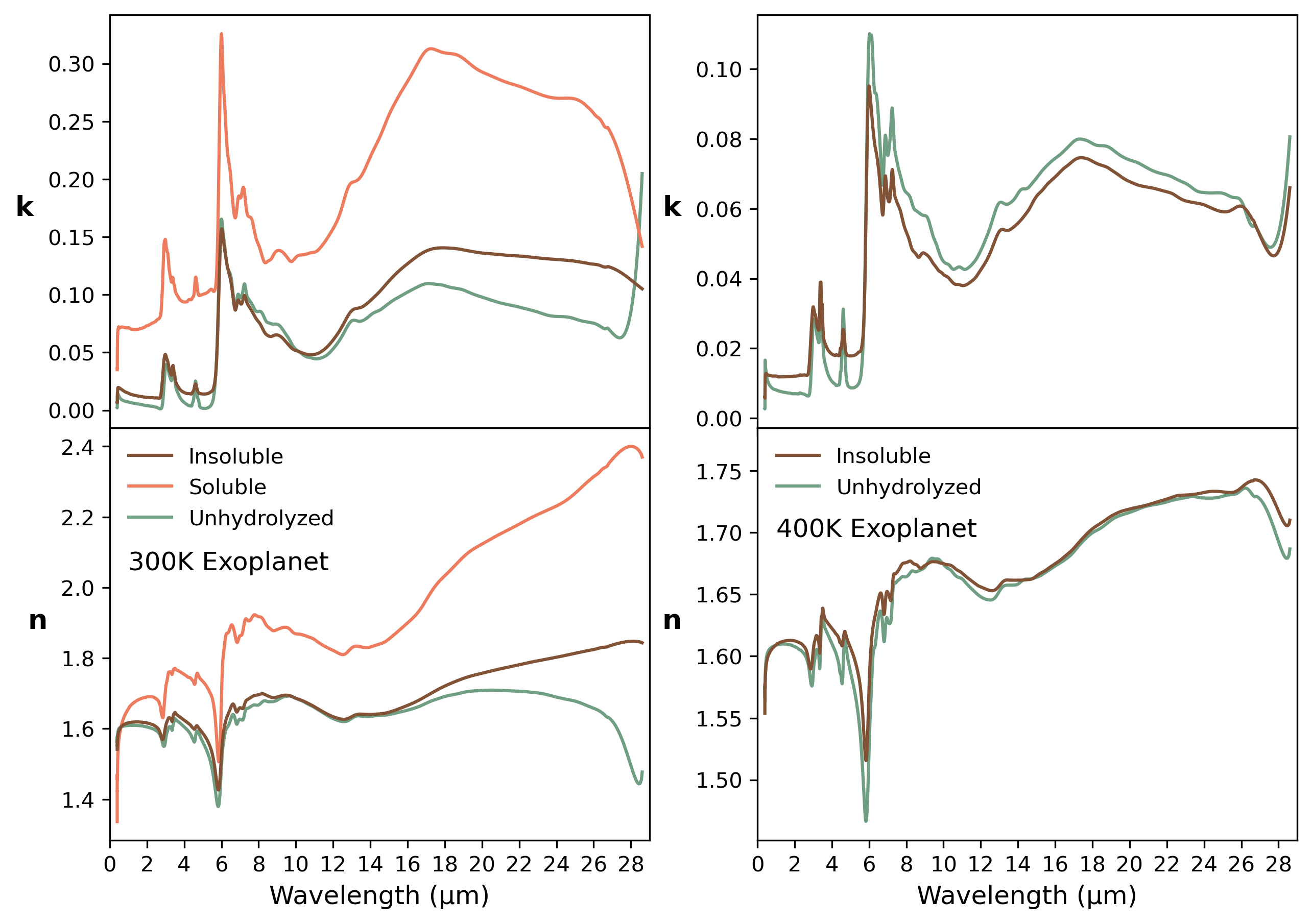}
    \caption{Optical constants (top: k, imaginary refractive indices; bottom: n, real refractive indices) of the insoluble and soluble portions of hydrolyzed hazes, along with the unhydrolyzed haze for comparison. Left: 300 K water-rich exoplanet; Right: 400 K water-rich exoplanet. The optical constants of hydrolyzed water-rich hazes are available as the Data behind the Figure.}
    \label{fig:optical}
\end{figure*}

Figure \ref{fig:transmittance} shows transmittance spectra of hydrolyzed and unhydrolyzed haze analogs for the simulated bodies. The spectral shape across hydrolysis states and samples are broadly similar. All spectra have an absorption feature at 0.42 $\mu$m, indicative of aromatic compounds and/or unsaturated species with conjugated pi bonds \citep{Rao1975,vanKrevelen2009}. The transmittance increases with wavelength from 0.5 to 2.5 $\mu$m, with differing abundances and structures of aromatic rings in the samples determining the degree of this slope as their absorptions compound to create a broad feature. It is important to note that the soluble fraction of the 300 K exoplanet was strongly diluted (to 20\% of the concentration of the other samples) to capture its features, meaning the sample is much more absorbing than the others (see Section \ref{sec:oc} and Fig. \ref{fig:optical}). The fingerprint mid-IR region from 2.5 to 20 $\mu$m introduces more specific spectral changes across our samples, exhibiting peak shifts, appearances, and disappearances that are indicative of organic functional groups evolving. The zoomed panels in Figure \ref{fig:transmittance} highlight this mid-IR region, and functional group assignments for the mid-IR are listed in Table \ref{tab:funcgroups}. 

The 300 K water-rich exoplanet analogs demonstrate functional group differences across all hydrolysis states. A free N-H feature appears at 3413 cm$^{-1}$ while a broad amine salt N-H feature from 3000--2800 cm$^{-1}$ is subsequently lost in the soluble sample. Because amine salts are readily soluble in water, this loss could reflect dissolution of the salts and release of free N-H groups. The persistence of the 3000--2800 cm$^{-1}$ feature in the insoluble fraction, however, suggests that either some salts remain undissolved or that another chemical mechanism contributes to the band. Varying intensities of the 2967, 2935, and 2877 cm$^{-1}$ C-H stretching modes may build into a broad feature in the insoluble material, consistent with the poor water solubility of nonpolar C-H bonds. 

Next, though not noted in Table \ref{tab:funcgroups}, the morphology of the broad 3500--2500 cm$^{-1}$ O-H stretching feature changes, with the hydrolyzed fractions starting to show this feature blueward of the unhydrolyzed. This is indicative of the O-H bonding structure changing from intramolecular to intermolecular bonding, or from bonding within the same molecule to bonding between separate molecules. As hydrolysis breaks apart large molecules, the exposed --OH groups on these new fragments are available to reassemble into aggregates via intermolecular bonds. Similar processes have been shown to assemble small, heterocyclic compounds into polymer-like chains resembling nucleotides without requiring intramolecular bonds \citep{Schuster2021}.

We next look to the 2300--2000 cm$^{-1}$ region of the spectra, associated with nitriles. The shift of the 2097 cm$^{-1}$ feature to $\sim$2140 cm$^{-1}$ in the soluble sample may indicate conversion of nitrile groups to carbodiimide species after hydrolysis, reorganizing the nitrogen-bearing bonds into more polar molecules. Moving to the $\sim$1600 cm$^{-1}$ region, where peptide backbones in amino acids present spectral features, we observe a loss of C=O in the insoluble fraction and N-H in the soluble fraction. These changes are consistent with amides breaking down into carboxylic acids and amines, producing reactive functional groups that may subsequently form larger, more complex molecules. The loss of one feature in the soluble fraction and the other in the insoluble likely reflects variations in bonding environments and solubility behaviors. 

Lastly, the C-H fingerprint region from approximately 1500-500 cm$^{-1}$ contains subtle information about the carbon bonding structure. All hydrolysis states contain features associated with both hydrocarbon chains and rings, however we detect the loss of a methylene group at 704 cm$^{-1}$ and appearance of one at 1340 cm$^{-1}$ in the soluble fraction. This suggests the transformation of some weakly substituted rings into methylene chains, though these vibrations are of limited diagnostic value without other structural measurements due to their strong coupling with other groups attached to the molecule.

The measured 300 K and 400 K samples share many similarities. While we might expect the soluble fractions to be similar as well, speculation on functional groups present in the 400 K soluble fraction is ineffective since the absence of soluble sample could reflect either the limited starting mass or compositional differences. Since most spectral changes across the 300 K samples occur in the soluble fraction, we lose information about the 400 K samples in the evaporated soluble fraction. In the measured 400 K analogs, we again observe the loss of methylene at 704 cm$^{-1}$. Additionally, the 1550 cm$^{-1}$ N-H feature is diminished in the insoluble fraction, whereas in the 300 K analog this difference appeared only in the soluble fraction. This adds to the complexity of deciphering the proposed hydrolysis reaction.

\subsection{Optical Constants of Hydrolyzed Haze}
\label{sec:oc}
Optical constants of hazes and other aerosols are imperative for creating model atmospheres and analyzing observational data. These values are fed to models to determine their effect on the radiative transfer of atmospheric gases, as well as determine whether their own spectral features can appear among the features of the gaseous background. The optical constants of hydrolyzed water-rich hazes are presented in Figure \ref{fig:optical}, with .txt files provided as the data behind this figure. 

For both haze analogs, the imaginary refractive indices of the insoluble fraction are similar in magnitude to the original haze analog. Values range from approximately 0.01 to 0.11. The imaginary refractive indices of the soluble fraction of the 300 K water-rich analog more than double at some wavelengths, spanning 0.05 to 0.30. This reflects the sample's altered chemistry that produces the high absorption seen in the transmittance spectra. These comparisons track to the real refractive index, where the insoluble and unhydrolyzed samples range from 1.4 to 1.8 and the soluble sample ranges from 1.5 to 2.4. A discussion on uncertainty in the optical constants can be found in Appendix \ref{sec:sensitivity}.

\begin{figure*}[th!]
    \centering
    \includegraphics[width=0.9\textwidth]{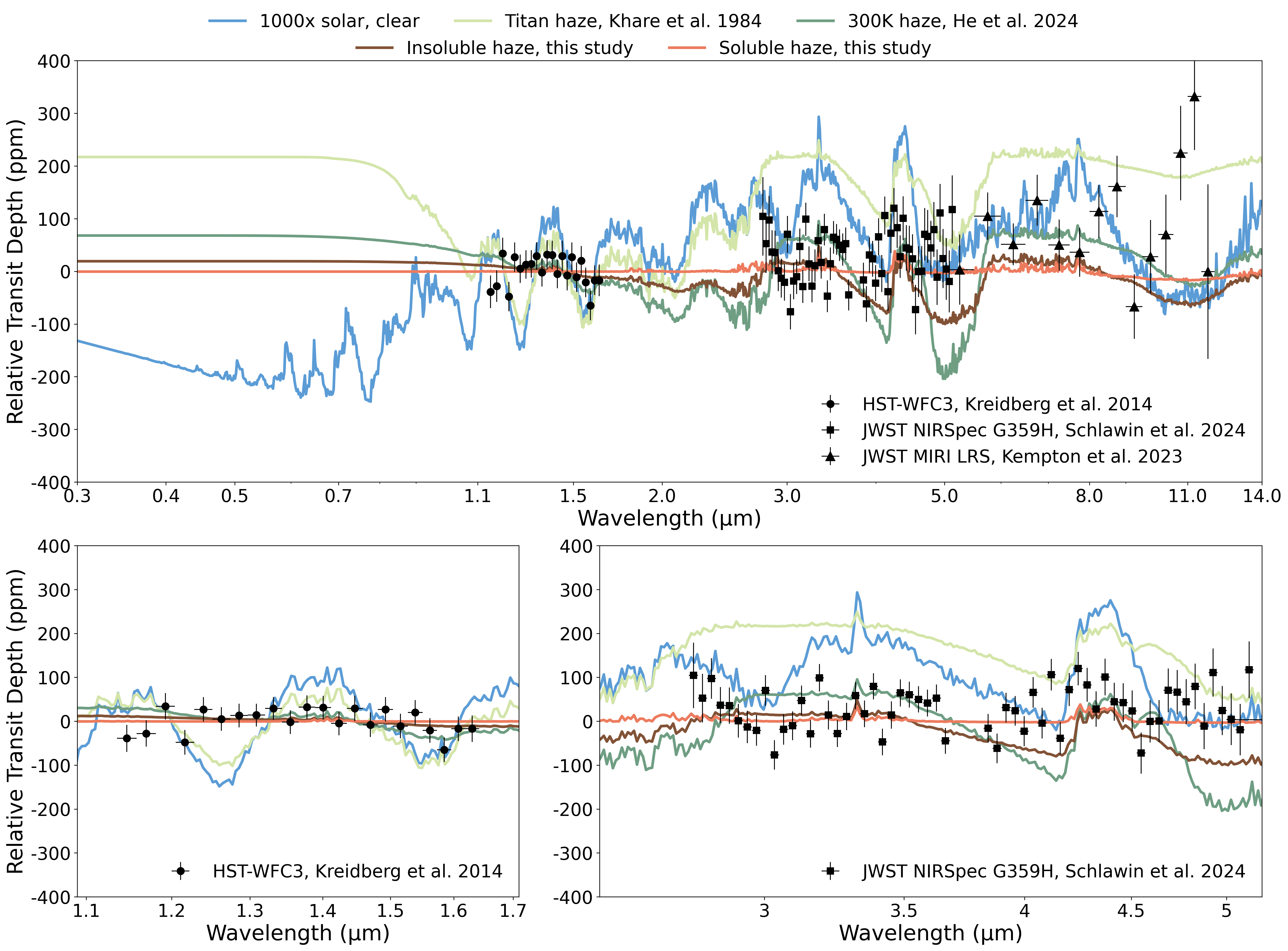}
    \caption{Model transmission spectra comparing several haze compositions on a GJ 1214b-like planet with a temperate water-rich atmosphere. We model a clear atmosphere as well as atmospheres with Titan-like hazes, water-rich exoplanet hazes, and the insoluble and soluble fractions of water-rich exoplanet hazes. Existing Hubble and JWST data of GJ 1214b are also plotted for reference. The top panel displays the full modeled wavelength range of 0.3 to 14.0 $\mu$m, and the bottom panels zoom into the wavelength regions with observational data.}
    \label{fig:transmission}
\end{figure*}

\begin{figure*}[th!]
    \centering
    \includegraphics[width=0.9\textwidth]{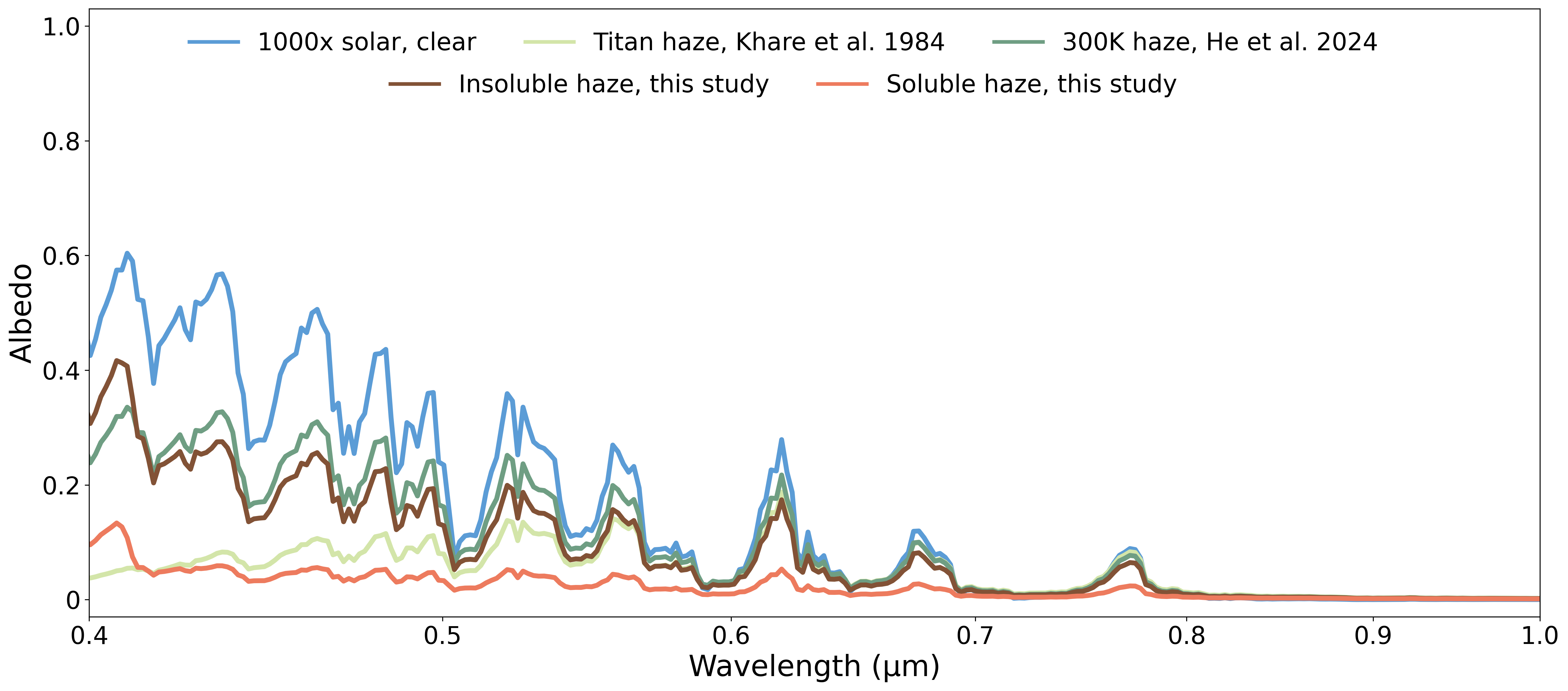}
    \includegraphics[width=0.9\textwidth]{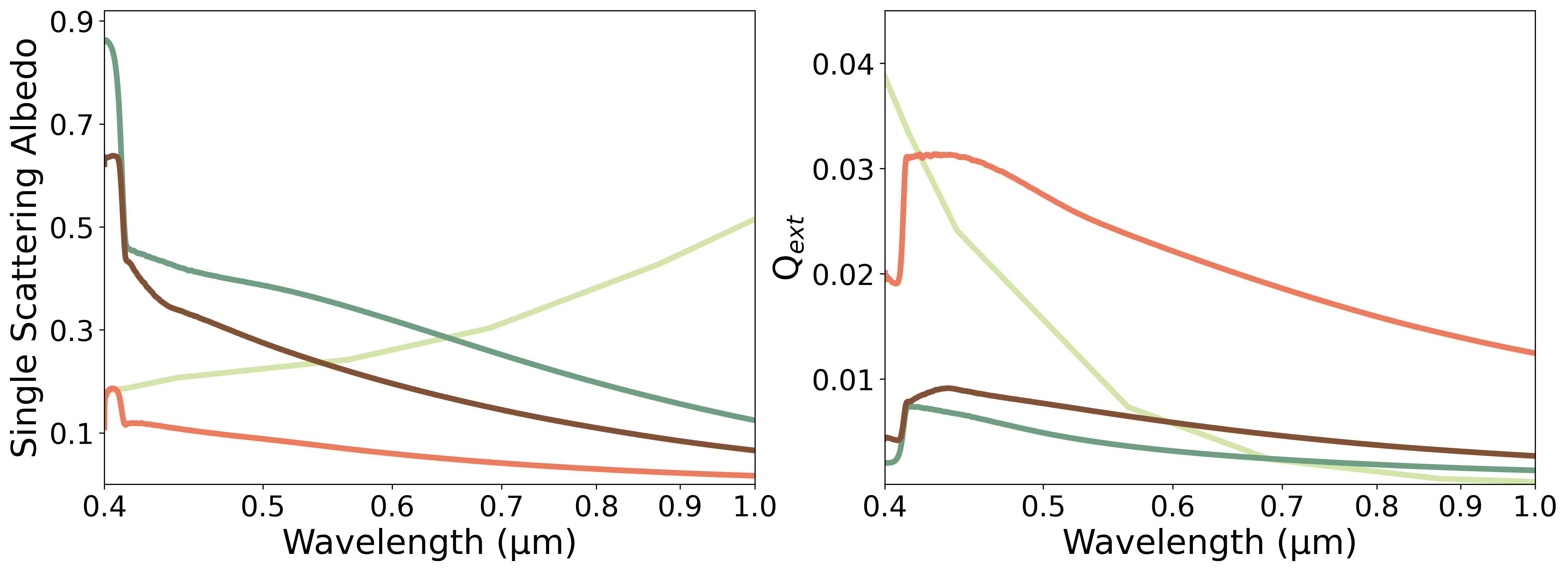}
    \caption{Top: Model reflectance spectra from 0.4 to 1.0 $\mu$m comparing several haze compositions on a GJ 1214b-like planet with a temperate water-rich atmosphere. We model a clear atmosphere as well as atmospheres with Titan-like hazes, water-rich exoplanet hazes, and the insoluble and soluble fractions of water-rich exoplanet hazes. Bottom: The main forcing model parameters (single-scattering albedo, left; extinction efficiency factor, right) for the haze cases. Scattering is summed over all of the radii in the particle size distribution used in the model. The Titan-like haze exhibits different trends than the various exoplanet hazes.}
    \label{fig:albedo}
\end{figure*}

\subsection{Hydrolyzed Hazes in Modeled Spectra}
To highlight the effect of hazes on observations, we present modeled transmission (Fig. \ref{fig:transmission}) and reflectance spectra (Fig. \ref{fig:albedo}) of a GJ 1214b-like planet with the same atmospheric composition as our 400 K water-rich atmosphere experiment (Table \ref{tab:mixingratio}). GJ 1214b is an ideal case study for aerosols in sub-Neptune atmospheres. Observational data from Hubble WFC3 G141 \citep{Kreidberg2014} and JWST \citep{Kempton2023,Schlawin2024} are plotted among the models in Figure \ref{fig:transmission} and demonstrate the need for a high metallicity atmosphere with high haze production \citep{Ohno2025}. All model parameters are held constant, including aerosol particle size and number density, only changing the optical constants of the injected hazes. We emphasize that the number densities of the hazes used in these models are higher than are expected to be physical, and the results here only demonstrate the importance of choosing optimal haze optical constants for any given atmosphere.

In transmission, the clear atmosphere has strong absorptions from CH$_4$, CO$_2$, and H$_2$O as well as a Rayleigh scattering slope in the optical. All hazy atmospheres mask the Rayleigh slope and many near-IR features, introducing a haze scattering slope, in addition to altering features in the mid-IR. There are prominent haze features seen at 3.0, 4.6, and 6.0 $\mu$m. Following previous discussion of the enhanced absorption in the soluble fraction of the hydrolyzed haze, spectral features are almost entirely muted in the soluble case, with the transit depth only reaching 55 ppm at the 4.3 $\mu$m CO$_2$ feature. The insoluble fraction is less muting though still a significant opacity source achieving a 130 ppm transit depth at 4.3 $\mu$m, whereas the Titan and original exoplanet hazes allow transit depths of 190 and 280 ppm at 4.3 $\mu$m, respectively. The Titan-like haze is the only scenario that allows for similarly large transit depths across the CH$_4$ and H$_2$O peaks in the 1.0 to 2.0 $\mu$m range.

\cite{Ohno2025} present comprehensive model-data comparisons from grid retrievals using several haze optical constants, looking for agreement with the featureless Hubble data and several weak absorptions in the NIRSpec G395H and MIRI LRS bands. The model transmission spectra are consistent with a high-metallicity atmosphere across haze properties, however specific features in the NIRSpec and MIRI bands vary based on the haze scenario. Adding the hazes produced in this work to the discussion in \cite{Ohno2025}, the hydrolyzed exoplanet haze cases show compelling agreement across datasets. Hydrolyzed haze models reproduce the featureless spectrum reported in \cite{Kreidberg2014} and, particularly for the insoluble fraction, may permit detection of the 4.3 $\mu$m CO$_2$ feature reported in \cite{Schlawin2024}. High absorption from the hydrolyzed hazes can also suppress the 3.0 and 4.6 $\mu$m haze features, matching the observed spectra without the need for such increased atmospheric metallicity as suggested by \cite{Ohno2025}. The data may align best with the insoluble fraction, which allows the 4.6 $\mu$m bump to emerge slightly as suggested by \cite{Ohno2025}. While the MIRI data have large error bars, the hydrolyzed hazes again eliminate the need for high metallicity atmospheres to achieve the statistically flat spectrum in \cite{Kempton2023}. Quantitative analysis is beyond the scope of this work, however robust statistical analysis would confirm the degree to which hydrolyzed hazes reproduce the observed trends.

Distinguishing between these haze scenarios in observations is currently limited by instrument capabilities and uncertainties. The signal-to-noise ratio for these spectral features is low, and offsets between the instrument detectors are unknown. As emphasized by \cite{Ohno2025}, follow-up observations with NIRSpec and MIRI may improve the signal-to-noise ratio enough to directly observe hazes in an exoplanet atmosphere for the first time.

The albedo spectra in Figure \ref{fig:albedo} reflect similar takeaways. The clear spectrum is dominated by H$_2$O absorptions until the gases become completely opaque around 0.84 $\mu$m. Each haze scenario {flattens the albedo spectrum}, though the haze chemistries affect spectra differently than in transmission. Here, the Titan-like hazes have a much larger dampening effect on spectral features than in transmission. The unhydrolyzed and insoluble haze particles behave similarly to each other, and the soluble exoplanet haze scenario still hosts the smallest features. These influences will be important for interpreting data from the upcoming Roman Space Telescope (0.5 to 0.8 $\mu$m) and Habitable Worlds Observatory (HWO; UV to infrared). Roman, a technological precursor to HWO, will provide important constraints on the observability of hazy worlds, where low albedo hazes may prove detrimental to observation even on the larger Jupiter analogs targeted by the telescope \citep{Bailey2023}. {HWO will target sub-Neptunes and other temperate worlds in the habitable zones of Sun-like stars that are not accessible to JWST transmission spectroscopy} \citep[e.g.,][]{Hu2025} and may be more sensitive to aerosol composition, particle size, and energy balance \citep{Gordon2025}.

In addition to supporting future telescopes, this wavelength region prompts an interesting discussion on wavelength-dependent optical properties. The Titan-like hazes studied in \cite{Khare1984} are famously bright \citep{Brasse2015}, having low absorption and high scattering across most of their broad measurement range (0.025 to 1000 $\mu$m). Based on this, we would expect this haze composition to exhibit the highest albedo {potentially even surpassing the clear atmosphere albedo}. By comparing the relevant parameters in Figure \ref{fig:albedo}, however, we see that the overarching trends in single-scattering albedo (SSA) and extinction efficiency factor do not hold from 0.4 to 1.0 $\mu$m. The Titan-like hazes from \cite{Khare1984} have a relatively low SSA and high extinction below 0.6 $\mu$m, where atmospheric gas opacity is still weak and our albedo spectra are dominated by haze opacity. The high reflectivity of Titan-like hazes only begins at wavelengths where atmospheric gases become the dominant influence in the spectra. Indeed, the albedo in the Titan-like haze case is largest, other than the clear case, in the 0.77 $\mu$m window after SSA increases and extinction decreases. Other laboratory Titan haze analogs disagree with the SSA and extinction efficiency in \cite{Khare1984}, varying based on experimental variables such as production temperature, pressure, and energy source \citep[e.g.,][]{Vuitton2009,Imanaka2012,He2022}. This again underscores the importance of choosing haze properties representative of the atmospheric conditions, as well as underscoring the necessity of laboratory work to explore numerous environmental conditions and cover broad wavelength ranges.

{We also note that the choice of model parameters is affecting the overall reflectivity of the atmosphere. It is somewhat surprising that all hazy scenarios decrease albedo from the clear scenario, in contrast to other reflected light aerosol studies \citep[e.g.,][]{Gao2017}, however a combination of haze optical properties, vertical extent, and mass loading in the model affect the radiative transfer. The large haze mass loading used in our model is spread over a large vertical extent; this decreased number density means that photons can travel deeper into the atmosphere before being scattered, experience longer path lengths, and have a higher probability of absorption before escaping. A shallower haze with similar mass loading would increase the likelihood of early scattering directly out of the atmosphere. This effect is compounded by the higher absorptivity of the organic hazes discussed above, demonstrating that the decreased albedo compared to clear is likely an effect of the material properties as well as model parameter choices.}

\subsection{Implications for Exoplanet Environments and Prospects for Observability}
On water-rich planets, water may be present in both the atmosphere and on the surface in the form of clouds, rain droplets, lakes, and oceans. This presents ample opportunity for photochemical hazes to meet liquid water. It has been hypothesized that hazes are good CCN, and the relatively high solubility of our haze analogs suggests that these hazes may be excellent CCN (see Section \ref{sec:solubility}; also \citealt{Moran2020}; \citealt{Yu2021}). Hydrolysis can occur at this interface inside cloud droplets, with our results showing that only a few weeks are required to alter haze chemistry. Time is an important factor, however, as previous studies of these haze analogs found much smaller solubilities during brief, hour-long solubility tests \citep{Yu2021}. Cloud lifetimes in planetary atmospheres are not well constrained, as examples from the solar system suggest that localized clouds may dissipate within hours to days \citep{Reuter2007,Brown2010} while global cloud decks may persist for months to years \citep{Titov2018,Sanchez2023}. Due to this variability, haze particles in a real system likely exist in both hydrolyzed and unaltered forms depending on the local conditions. The partitioning between soluble and insoluble material in a real atmosphere will also be complex due to reaction times and volatile evaporation conditions. Comprehensive modeling studies should account for mixtures of haze hydrolysis states. 

As clouds dissipate, the solid products could remain suspended in the atmosphere or be transported to lower altitudes via rainfall, and the volatile hydrolysis products may evaporate along with water. With such a large fraction of the mass becoming volatile, these products would not simply disappear. Instead, they may themselves be condensable and form additional cloud layers, further impacting observations beyond refractory hydrolysis residues. HCN on Titan, for example, is a common gas-phase photochemical product that condenses in colder regions of the atmosphere \citep{Lavvas2011}. Similar processes may occur for common low molecular weight hydrolysis products such as ammonia, formamide, and small amines. The extent of this process is highly uncertain, as it depends on the species' vapor pressures in addition to the temperature-pressure profile of the atmosphere.

Hazes can also bypass chemistry in cloud layers by settling into deeper atmospheric layers through dry deposition. This process can be compared to Titan, where haze particles are thought to supply enough material to make up the majority of the moon's surface \citep[e.g.,][]{Cable2012}, meeting transient melt pools formed by cryovolcanism or impacts that may last for $\sim$10$^2$--10$^4$ years \citep{Neish2006}. While sub-Neptunes are unlikely to have a well-defined surface, they are hypothesized to host liquid or supercritical water ocean layers beneath the cloud deck \citep{Nixon2021,Luu2024}. The water-rich layers predicted for sub-Neptunes are dynamically coupled to their atmospheres \citep{Morley2015,Charnay2015}, enabling exchange between haze-forming regions and deep liquid reservoirs. The conditions of our experiments are not representative of deep interiors, however efficient exchange means that hazes can not only be chemically altered by the deep environment but can also be lofted back up into observable atmospheric layers through convective processes. 

Given these formation and transportation processes, hydrolyzed hazes may subsist at observable altitudes. The altitude of water clouds is highly dependent on an atmosphere's water mixing ratio, temperature, and pressure, but the presence of hazes can change where clouds are stable. With haze as CCN, condensation can occur at higher altitudes or lower water mixing ratios, shifting the cloud deck into regions clearly observable in transmission and reflectance. If hazes settle deeper into the atmosphere, observations may instead probe lower atmospheric layers through transparent windows of the atmospheric gases, as has been done to study the surface of Titan by gazing through CH$_4$ absorptions \citep[e.g.,][]{McCord2006}. Conversely, the vigorous vertical mixing expected in these atmospheres can bring hydrolyzed hazes back to higher altitudes. Together, hydrolyzed hazes are within observable reach of many state-of-the-art telescopes including JWST, Roman, and HWO.

\subsection{Future Work}
Avenues for future work exist both in the laboratory and out. In the laboratory, chemical characterization of the liquid hydrolysis products will provide a fuller picture of volatile species in the atmosphere and prebiotic chemistry in aqueous environments. For the solid products, mass spectrometry techniques may be employed to determine possible molecular structures and look for prebiotic molecules. These experiments would also benefit from developing a thin film hydrolysis method, rather than the powder hydrolysis method employed in this study. With a thin film, the photochemical product is deposited directly onto a substrate, rather than being collected with spatulas. This preserves the particle morphology to enable measurements of physical properties such as particle size and surface energy. These properties tell us about particle lifetimes, cloud formation, and scattering properties. Developing this method would also enable reflectance spectroscopy, improving the optical constants derivation.

Outside of the laboratory, the results presented here can be incorporated into models and observations. Cloud microphysical models can take haze solubility into account using the K{\"o}hler Equation to better predict cloud coverage. Atmospheric dynamics and radiative transfer models can explore how hydrolyzed hazes may affect a planet's energy budget. Lastly, synthetic atmospheric spectra can utilize more specific planetary parameters and robust statistical analysis to make quantitative comparisons to observations. There is much still to be explored regarding hazes in water-rich atmospheres.

\section{Conclusions}
This experimental study explores the interaction of water with photochemical hazes in sub-Neptune atmospheres. Haze analogs for 300 and 400 K water-rich exoplanet atmospheres have relatively high solubilities, influencing cloud formation and altering optical properties. There is spectral evidence of changes in key functional groups after hydrolysis involving N-H, O-H, C$\equiv$N, C=O, and C-H bonds. Due to these differences, atmospheric models and retrievals must use representative optical constants for the atmosphere in question. Our results demonstrate that hazes formed in and weathered by a water-rich environment can greatly flatten spectral features, and broadband wavelength measurements are required to fully understand the spectral implications. Future work should continue to aid observations by uncovering the properties of haze particles. These studies will begin to illuminate what lies underneath hazy exoplanet atmospheres.

\section{Acknowledgments}
This work was supported by the NASA Exoplanets
Research Program grant NNX16AB45G and the NSF Astronomy and Astrophysics Grant Program grant 2206245. S.E.M. is supported by NASA through the NASA Hubble Fellowship grant HST-HF2-51563 awarded by the Space Telescope Science Institute, which is operated by the Association of Universities for Research in Astronomy, Inc., for NASA, under contract NAS5-26555.

\clearpage
\bibliography{mybib}

\appendix

\section{Sensitivity to Assumptions and Estimated Uncertainties}
\label{sec:sensitivity}

\begin{figure*}[th!]
    \centering
    \includegraphics[width=0.3\textwidth]{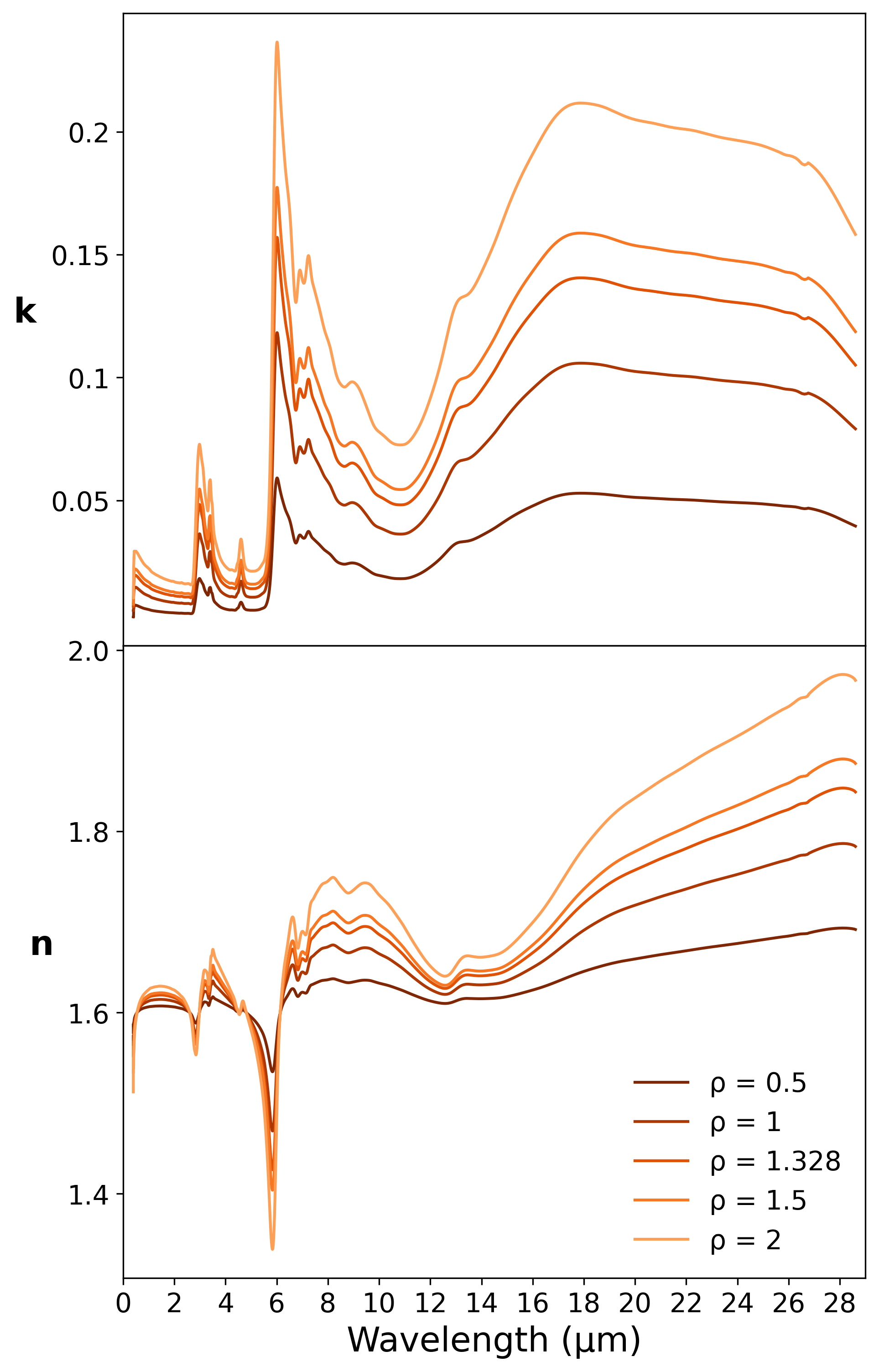}
    \raisebox{0.5\height}{\includegraphics[width=0.3\textwidth]{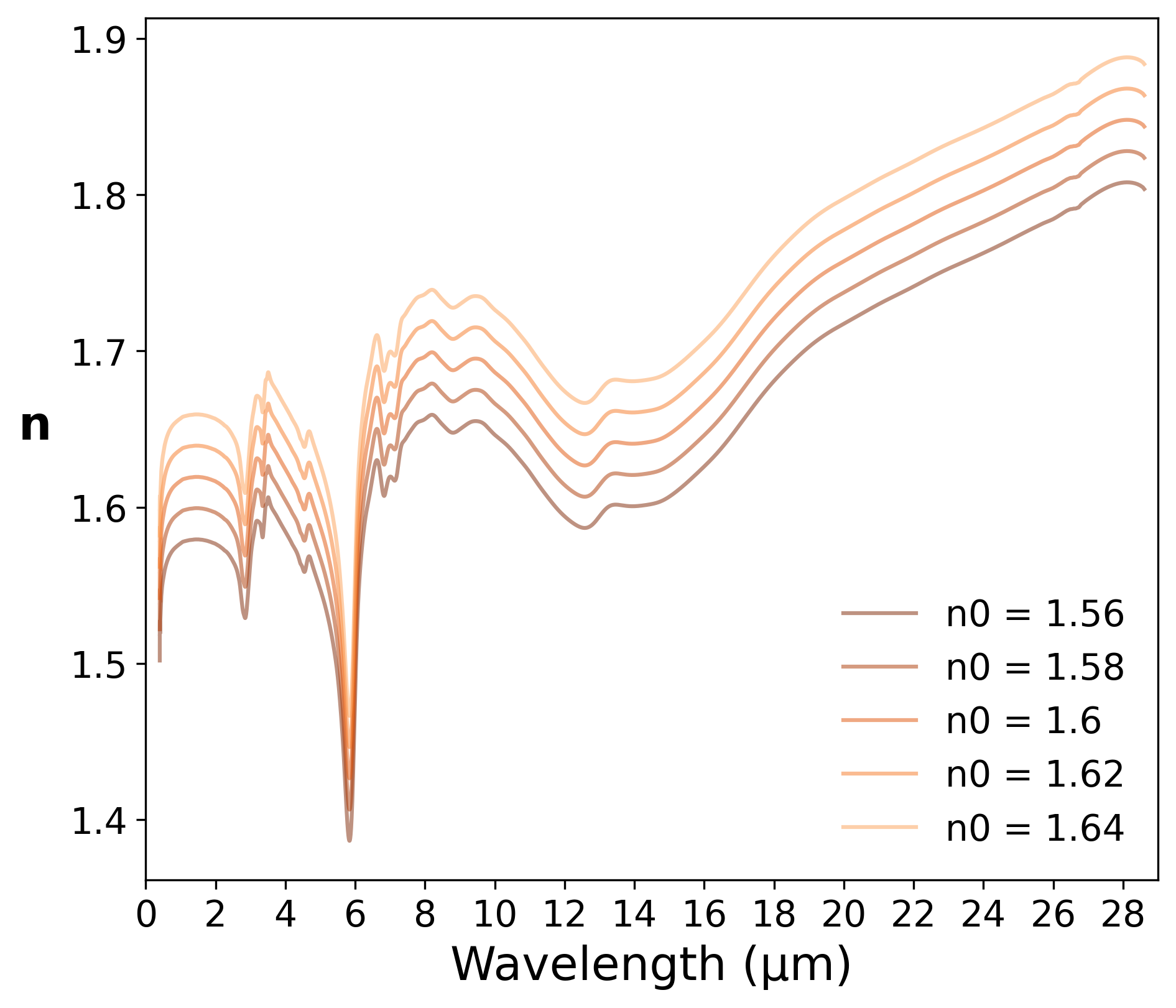}}
    \raisebox{0.5\height}{\includegraphics[width=0.3\textwidth]{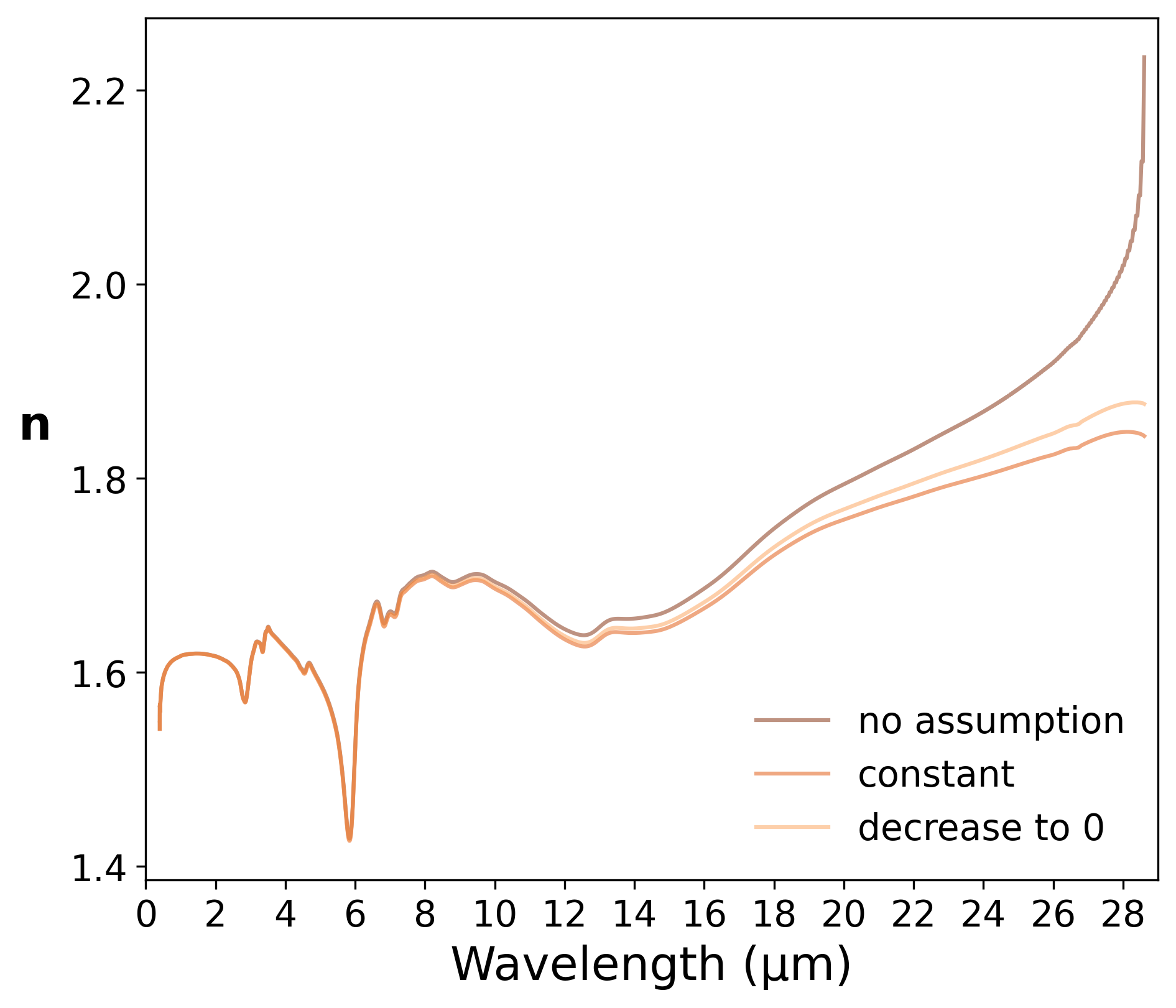}}
    \caption{Optical constants for the insoluble fraction of the 300 K hydrolyzed haze analog derived under various assumptions. Left: $n$ and $k$ derived from particle densities ($\rho$) spanning 0.5 to 2 g cm$^{-3}$. Middle: $n$ derived using anchor points $n_0$ from 1.56 to 1.64. Right: $n$ derived from differing assumptions for $k$ outside of the measured wavelength range.}
    \label{fig:sensitivity}
\end{figure*}

The optical constants derivation employed in this study makes several assumptions regarding sample properties that are unable to be verified experimentally. Here, we discuss the validity of these assumptions and test alternate choices. The uncertainty of our optical constants factor in these sensitivity tests in addition to instrument uncertainty. 

Sample density is used to determine the effective thickness of the KBr pellet (Eq. \ref{eq:d}). Density can be measured using a gas pycnometer, however we do not recover close to the $\sim$90 mg of sample required for this method. Instead, we assume the same density as the unhydrolyzed haze analogs (300 K: 1.328 g cm$^{-3}$; 400 K: 1.262 g cm$^{-3}$). In reality, the density of the hydrolyzed sample likely increases due to changes in oxygen incorporation, molecular weight, polarity, and degree of unsaturation. \cite{He2017} find that the densities of Titan haze analogs increase by 6\% from an anoxic initial gas mixture to one with 5\% CO. An increase of 6\% would increase the exoplanet haze analog densities to 1.407 and 1.338 g cm$^{-3}$ for the 300 and 400 K cases, respectively, matching well with various Titan haze analog measurements \citep{Imanaka2012,He2017}. It is of course possible that the density increases by more than 6\%, and we must also consider that some Titan haze analog measurements give much lower densities, from 0.5 to 1.1 g cm$^{-3}$ \citep{Trainer2006,Horst2013}. A large range of densities is worth exploring, as 0.5 to 2 g cm$^{-3}$ are all reasonable densities for organic solids. The left panel in Figure \ref{fig:sensitivity} shows representative optical constants propagated with densities varying from 0.5 to 2.0 g cm$^{-3}$. The extinction coefficient $k$ is inversely related to density, so the effect is quite large at a 62\% deviation from the assumed density. The real refractive index $n$ is far less affected, as we integrate $k$ over all wavelengths. Importantly, the extremes of the studied densities change the overall transit depth of the synthetic spectra ($\sim$14,400 ppm) by less than 1\%. For instance, the 4.3 $\mu$m CO$_2$ feature amplitude ranges from 90 to 200 ppm across the tested densities, compared to 130 ppm for our baseline assumption. While measuring the haze density would improve confidence in distinguishing between haze scenarios, the key takeaways of this work are not expected to change.

We also assume an anchor point for $n$ ($n_0$), meant to reduce uncertainty in the numerical integration in the SKK relation (Eq. \ref{eq:rri}). To obtain this value, we need another method of optical constant derivation. The method compatible with our experimental setup uses reflectance spectroscopy of a thin film of sample to compare interference fringes at two angles. For a typical haze analog, thin films are produced during sample production as material builds up on a substrate in a uniform layer. This procedure is complicated by the post-processing involved in hydrolysis. Simply submerging an unhydrolyzed thin film in water poses two issues: we cannot assume that all layers of the film are exposed to water, and we cannot separate soluble and insoluble fractions. As a result, the reflectance spectra would probe a mixture of unhydrolyzed, soluble, and insoluble material, making interpretation difficult. We also explored drying a small amount of hydrolyzed powder onto a substrate, however we cannot guarantee a uniform thickness and therefore cannot accurately compare measurements from two angles. Given that most organic polymers have $n$ values between 1.5 and 1.6 \citep{Zhang2023}, we can test various reasonable $n_0$ values. We are also constrained by the $n_0$ of the original samples, determined to be 1.6027 for the 300 K sample and 1.6213 for the 400 K sample \citep{He2024}. The middle panel in Figure \ref{fig:sensitivity} compares $n$ derived from $n_0$ between 1.56 and 1.64. All $n$ curves vary from the chosen anchor point of 1.6 by less than 2.5\%, and the resulting synthetic atmospheric spectra change by less than 0.1\%. The 4.3 $\mu$m CO$_2$ feature amplitude changes by less than 10 ppm. Our results are very robust to the choice of $n_0$.

The last assumption factoring into the optical constants derivation is the assumed value of $k$ outside of the measured wavelength range, required in the integration in Equation \ref{eq:rri}. Generally, assuming a constant $k$ is valid unless there are large local absorptions beyond the measurement range. Nonetheless, we test several cases to understand the effect that this choice has on $n$. The right panel in Figure \ref{fig:sensitivity} shows $n$ for three end-member cases: assuming a constant $k$ equal to the last $k$ value on either end of the measured wavelength range, assuming a monotonic decrease to 0, and not integrating past the measured wavelength range. We see the most sensitivity in the mid-IR, peaking at 18\% for the derivation that does not integrate outside the measured range, though the vast majority of the deviation is $<$3\%. Synthetic atmospheric spectra are again affected by less than 1\%, leading to less than 10 ppm changes in the 4.3 $\mu$m CO$_2$ feature amplitude. Our results are also robust to the handling of $k$ outside the measured wavelengths.

\end{document}